\documentclass[aps,prc,twocolumn,showpacs,tightenlines,superscriptaddress,floatfix]{revtex4}
\usepackage{braket}
\usepackage{graphicx}
\usepackage{amssymb,amsmath}
\usepackage[]{color}

\def\vec#1{{\boldsymbol{#1}}}
\newcommand{\nuc}[2]{\hbox{$^{#1}$#2}}

\newcommand{\s}{\mathcal{S}}
\newcommand{\ms}[1]{|\mathcal{S}_{#1}|^2}
\newcommand{\hr}{\hat{\vec{r}}}
\newcommand{\vecr}{\vec{r}}
\newcommand{\hl}{\hat{\ell}}

\newcommand{\acf}{\Gamma_{\ell_1\ell_2\ell_1'\ell_2'}^{L}(\omega)}
\newcommand{\acfx}{\Gamma_{\ell_1\ell_2\ell_2'\ell_1'}^{L}(\omega)}

\newcommand{\cg}[1]{(#1)}

\newcommand{\amp}{\mathfrak{C}_{\alpha{LS}}^{IT}}
\newcommand{\ampp}{\mathfrak{C}_{\alpha'{LS}}^{IT}}

\newcommand{\cfc}[1]{\mathcal{C}_{\ell_{#1}\lambda_{#1}}}

\newcommand{\ray}[2]{\mathcal{R}_{\beta_{#1}}^{\lambda_{#1}}\!(#2)\,}
\newcommand{\rayp}[2]{\mathcal{R}_{\beta_{#1}'}^{\lambda_{#1}'}\!(#2)^*\,}
\newcommand{\hay}[3]{\mathcal{H}_{\lambda_{#1}\lambda_{#2}'}(#3)}
\newcommand{\pwt}[2]{\psi_{\beta_{#1}}^{\lambda_{#1}}(\vecr_{#2})}
\newcommand{\pwpt}[2]{\psi_{\beta_{#1}'}^{\lambda_{#1}'}(\vecr_{#2})}

\begin{document}

\title{Correlations probed in direct two-nucleon removal reactions}

\author{E.\,C. Simpson}
    \affiliation{Department of Physics, Faculty of Engineering and
      Physical Sciences, University of Surrey, Guildford,
      Surrey GU2 7XH, United Kingdom}
\author{J.\,A. Tostevin}
    \affiliation{Department of Physics, Faculty of Engineering and
      Physical Sciences, University of Surrey, Guildford,
      Surrey GU2 7XH, United Kingdom}

\date{\today}
\begin{abstract}
Final-state-exclusive momentum distributions of fast, forward
travelling residual nuclei, following two nucleon removal from fast
secondary radioactive beams of projectile nuclei, can and have now
been measured. Assuming that the most important reaction mechanism
is the sudden direct removal of a pair of nucleons from a set of
relatively simple, active shell-model orbital configurations, such
distributions were predicted to depend strongly on the total angular
momentum $I$ carried by the two nucleons -- the final state spin for
spin $0^+$ projectiles. The sensitivity of these now-accessible
observables to specific details of the (correlated) two-nucleon wave
functions is of importance. We clarify that it is the total orbital
angular momentum $L$ of the two nucleons that is the primary factor
in determining the shapes and widths of the calculated momentum
distributions. It follows that, with accurate measurements, this
dependence upon the $L$ make-up of the two-nucleon wave functions
could be used to assess the accuracy of (shell- or many-body) model
predictions of these two-nucleon configurations. By use of several
tailored examples, with specific combinations of active two-nucleon
orbitals, we demonstrate that more subtle structure aspects may be
observed, allowing such reactions to probe and/or confirm the
details of theoretical model wave functions.
\end{abstract}
\maketitle

\section{Introduction}

The momentum distributions of the residual nuclei, following the
removal of a single nucleon from a fast radioactive secondary beam,
offer sensitive probes of both strongly-bound and weakly-bound
single-particle structure near the (asymmetric) Fermi surfaces of
neutron-rich and neutron-deficient nuclei. Specifically, the shapes
and widths of the exclusive residue momentum distributions were
shown to be characteristic of the orbital angular momentum of the
removed nucleon \cite{Han96, BaV98,HaT03,BeH04,Gad08}.

The simplest generalization to the case of direct two-nucleon
removal is to describe the wave function of the two nucleons in the
projectile by a product of nucleon wave functions in assumed
single-particle orbitals. Doing so, the two nucleons are
uncorrelated, other than both being bound to the same core
\cite{BBC03,jatrnb7}. The heavy residue longitudinal momentum
distributions in this limit, being essentially the convolution of
those of the single-nucleons, depend on the assumed quantum numbers
of the two nucleons, but, in the absence of explicit
antisymmetrization or total angular momentum coupling of the two
nucleons, are not characteristic of specific residue final states
\cite{jatrnb7}.

More recent theoretical developments now treat fully the shell-model
correlations of the two removed nucleons in the projectile many-body
wave function \cite{TPB04,ToB06}. In the fully-correlated models the
product of nucleonic wave functions is replaced by the shell-model
two-nucleon overlap, incorporating (i) the two-nucleon parentage
coefficients with respect to each residue final state (the two
nucleon amplitudes, or TNA), (ii) proper antisymmetrization of the
two removed nucleons, and (iii) proper angular momentum coupling.

The resulting theoretical description, and the insights developed here, 
are equally valid for reactions that remove two loosely- or strongly-
bound nucleons. However, as has been discussed elsewhere \cite{TPB04}, 
in the case of removal of two loosely-bound nucleons the direct removal 
cross sections, of interest here, will be overwhelmed (experimentally) 
by indirect reaction (one-nucleon removal plus evaporation) events. See 
for example reference \cite{SiT09} for a quantitative
consideration 
of the direct and indirect two-neutron removal reaction contributions 
along the neutron-rich carbon isotopic chain. For these reasons we will
restrict our attention to examples for which the removed nucleons are 
strongly-bound, where the indirect removal paths are effectively closed,
and for which the direct cross sections are accessible experimentally.

Demonstrative test cases, e.g. in Ref. \cite{STB09b}, assumed the
two nucleons originated from a single orbital -- a pure
configuration. In this limit the TNA enter only as a multiplicative
(spectroscopic-like) factor and thus the new and interesting
characteristics of the residue momentum distribution are a result of
the correlations due to antisymmetrization and angular momentum
coupling.

These developments demonstrate the potential of two-nucleon removal
for exotic nucleus spectroscopy, showing the final-state-exclusive
residue nucleus momentum distributions to have shapes and widths
that are characteristic of the total angular momentum, $I$, carried
by the removed pair of nucleons -- and permitting final state spin
assignments to be made \cite{STB09a}. For the spin $J_i=0^+$
projectile nucleus examples used in Ref. \cite{STB09b}, there was a
high sensitivity of the residue momentum distributions to the final
state total angular momentum $J_f=I$. Moreover, the shapes of these
calculated distributions were robust with respect to variations
of other key structure and reaction parameters, such as the nucleon
separation energy. Further, the consideration of pure configuration
examples (e.g. two protons, assumed removed from a single active
$\pi 1d_{5/2}$, $\pi 1d_{3/2}$ or $\pi 2s_{1/2}$ orbital) showed
considerable insensitivity of the two-nucleon removal distributions
to these individual nucleon quantum numbers; in stark contrast to
results from single-nucleon removal reactions where the orbital
angular momentum is critical.

Thus, although the two-nucleon removal process is powerful for
final-state spin spectroscopy in very exotic systems, its
sensitivity to and ability to probe finer details of the shell-model
wave functions and the two nucleon configurations and correlations
therein remains less clear. Already in Fig. 5 of Ref.
\cite{jatrnb7}, for the case of two-proton removal from
\nuc{28}{Mg}, using the fully-correlated shell-model wave functions
for the transitions to the first two low-lying \nuc{26}{Ne}($2^+)$
final states one observed momentum distributions with
different widths; demonstrating a sensitivity \emph{beyond} the
final state spin. Our objective here is to elucidate this
sensitivity of the calculated residue momentum distributions to the
particular two-nucleon configurations present and to understand the
sensitivity to the combination of orbitals involved for a given
value of the pair's total angular momentum $I$.

In contrast to two-nucleon transfer reactions, such as the ($p,^3$H)
reaction, wherein the $\langle p|^3 $H$ \rangle$ light-ion structure
vertex preferentially selects the pick-up of a spin-singlet ($S=0$)
neutron pair, the two-nucleon removal mechanism is not explicitly
selective in the nucleon spins \cite{TPB04}. Both spin-singlet and
spin-triplet components of the two-nucleon overlap will be probed
and, under the assumption that the residue and nucleon-target
interactions (and $S$-matrices) are spin-independent, the $S=0$ and
$S=1$ terms contribute incoherently to the reaction yield. We will
show specifically, in the same way that single-nucleon removal is
sensitive principally to the orbital angular momentum $\ell$ rather
than the total angular momentum $j$ of the nucleon, that the
two-nucleon removal reaction momentum distributions are sensitive to
the components in the two-nucleon overlap with a given value of
total orbital angular momentum $L$, $\vec{L}=\vec{\ell}_1
+\vec{\ell}_2$. The presence and relative strengths of these $L$
components are determined via the shell-model overlaps and their
TNA. Since, for a spin-zero projectile, the residue has total spin
$J_f$, with $\vec{J}_f=\vec{L}+\vec{S}$, the total spin content of
the overlap is determined by the nuclear structure. However, all
spin components present are sampled by the reaction mechanism.

Recognizing the sensitivity to $L$ allows greater probing of the
shell-model wave function, particularly in states where the mixing
of several available nucleon configurations may be weak. Specific
examples can exhibit particularly strong sensitivity to the orbital
combinations of the pair. In cases of strong mixing and sharing of
strength among several orbitals a particular value of $L$ may
nevertheless be favored. In such cases the residue momentum
distribution will be characteristic of the details of the states
populated and not simply of the total angular momentum value $I$ of
the nucleon pair.
Of particular interest will be different regions of $A$ and $Z$ that
affect the originating, active orbitals of the two nucleons.

In Section \ref{sec:formalism} we outline the approximations assumed
and then present the required formalism for residue momentum
distributions within the $LS$-coupling scheme. We will retain
isospin labels for clarification of the underlying symmetries. In
Section \ref{corr_sect} the importance of the total orbital angular
momentum is elucidated by a detailed consideration of the spatial
and angular correlations of the two nucleons that are inherent in
the two-nucleon overlap function. This analysis also provides
insight into the possible sensitivity to the mixing of orbitals
across major shells. In Section \ref{examp_sect} we then consider
particular examples with projectiles of different $A$ and $Z$ where
interesting effects are predicted. Examples will look at specific
final states that could be populated in two-nucleon removal from the
$p$-shell, \nuc{12}{C}($-np)$, the $sd$-shell, \nuc{26}{Si}($-2n)$,
and also the $sdpf$-cross shell situation in \nuc{54}{Ti}($-2p)$. We
summarize the article and draw conclusions in the final Section.

\section{Formalism\label{sec:formalism}}

We discuss the sudden, direct removal of two nucleons from a fast
projectile beam incident on a light nuclear target at energies of
order 80 MeV per nucleon and greater.  
In this intermediate energy range there have been extensive (positive) 
assessments of the validity and accuracy of the sudden/adiabatic and eikonal 
reaction dynamical approximations (see e.g. Section 3.5 of Ref. \cite{arnps} 
and references therein) and of the theoretical ingredients used in their 
implementation; such as the importance of Pauli-blocking on the effective 
nucleon-nucleon interaction used \cite{Bertulani}.  These wide-ranging 
assessments, carried out within the one-nucleon stripping and breakup 
contexts, remain valid in the present analysis; e.g. the role of strong 
absorption between the projectile and target in reducing the effective reaction 
time and the energy for validity of the sudden/fast adiabatic approximation 
\cite{Summers}.

We first briefly discuss the
salient features of the approach, previously developed in detail in
Refs. \cite{TPB04,ToB06,STB09b}. Our emphasis here will be on the
use of the $LS$-coupling representation to describe the two-nucleon
structure overlaps and to derive the expressions for the residue
momentum distributions in this basis. We will consider only those
(stripping) reaction mechanism events in which the two removed
nucleons interact inelastically with the target nucleus. The role on
the cross sections and momentum distributions of the other major
class of events (diffraction-stripping), where one of the nucleons
is removed by an elastic interaction with the target, were discussed
fully in Refs. \cite{ToB06} and \cite{STB09b}, respectively. As was
shown there, events from this second mechanism give residue momenta
that are essentially identical to those of the stripping mechanism.
These conclusions remain unchanged and will not be repeated here.

We assume the projectile nucleus to be an antisymmetrized $A+2$-body
(shell-model) system, denoted by $\Psi_i \equiv \Psi_{J_i M_iT_i
\tau_i} (A,1,2)$, carrying total angular momentum $J_i$ and isospin
$T_i$. In a high-speed collision with the light target, two nucleons
may be removed to produce an $A$-body reaction residue in a final
state $f$, often referred to as the core state for simplicity. This
final core state is $\Phi^{(F)} \equiv \Phi_{J_fM_fT_f\tau_f}(A)$.
Each residue final state is denoted by $f$, while $F\equiv (f,M_f)$
is used to refer to a state with a specific angular momentum
projection $M_f$. One is reminded that the final state of the two
nucleons and of the target nucleus are unobserved and that the
observables discussed are inclusive with respect to these degrees of
freedom.

\subsection{Two-nucleon overlap \label{sec:overlap}}

The direct reaction will probe the two-nucleon overlap
\begin{align}
\Psi_i^{(F)} \equiv& \Psi_{J_i M_i T_i \tau_i}^{(F)}(1,2)\nonumber
\\ \equiv & \braket{\Phi^{(F)}(A)
|\Psi_i (A,1,2)} \nonumber\\ =&\sum_{I\mu{T}\alpha}C_{\alpha}^{IT}
\cg{I{\mu}J_fM_f|J_iM_i} \nonumber\\
&\cg{T{\tau}T_f\tau_f|T_i\tau_i} \;
[\,\overline{\psi_{\beta_1}(1)\otimes \psi_{\beta_2}(2)}\,]_{I
\mu}^{T\tau}\ , \label{eqn:sm_overlap_ls}
\end{align}
where $\Psi_i$ and $\Phi^{(F)}$ were defined above. The signed
two-nucleon amplitudes (TNA) $C_{\alpha}^{IT} {\equiv}
C_{\alpha}^{J_iJ_f IT_iT_fT}$ will be taken from shell-model
calculations. They express the parentage (amplitudes) for finding
each two-nucleon configuration $\alpha$ and residue final state $f$
in the overlap with the projectile initial state $i$, assumed to be
the ground state. The two-nucleon configurations index, $\alpha
\equiv [\beta_1,\beta_2]$, denotes the spherical quantum numbers of
the single-particle states occupied by the nucleon pair. Hence,
$\beta \equiv n\ell j$. Note that the amplitudes $C_{\alpha}^{IT}$
refer to a specific $i\rightarrow f$, initial to final state
transition, which will be understood implicitly.

The details of the shell-model calculations (e.g. the model spaces
and interactions) used to construct these overlaps will be presented
with each relevant example in the later Sections.

\begin{figure}[b]
\includegraphics[width=\columnwidth]{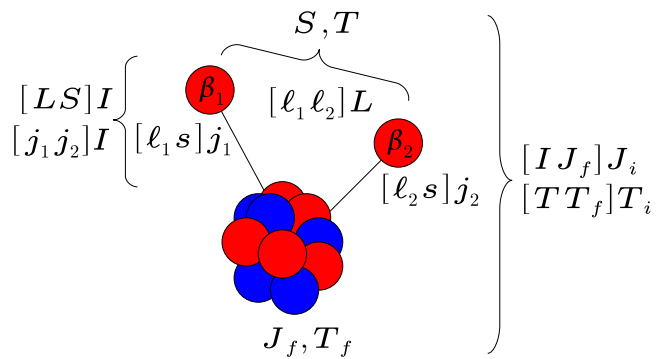}
\caption{Angular momentum and isospin coupling schemes used in the
calculations. The projectile initial (ground) state and final
residue state have spin $J_i$ (projection $M_i)$ and $J_f$
(projection $M_f)$ and isospins $T_i$ (projection $\tau_i)$ and
$T_f$ (projection $\tau_f)$, respectively. Each two nucleon
configuration $\alpha$ involves of a pair of active orbitals
$\beta_1$ and $\beta_2$ with spherical shell-model quantum numbers
$n_i$, $\ell_i$ and $j_i$ (projections $\lambda_i$ and $m_i)$. In
$LS$-coupling, the nucleon orbital angular momenta $\ell_1$ and
$\ell_2$ are coupled to $L$ (projection $\Lambda)$, the intrinsic
spins to $S$ (projection $\Sigma)$ and $L$ and $S$ to a total
angular momentum $I$, which must also couple the initial and final
state total angular momenta. \label{fig:coupling}}
\end{figure}

Expressed in $LS$-coupling, the antisymmetrized two-nucleon wave
function in Eq. (\ref{eqn:sm_overlap_ls}) is
\begin{widetext}
\begin{align}
[\,\overline{\psi_{\beta_1}(1)\otimes\psi_{\beta_2}(2)}\,]_{I\mu}^{T\tau}
=&
D_{\alpha}\hat{j}_1\hat{j}_2 \sum_{\substack{L\Lambda{S\Sigma}\\
\lambda_1\lambda_2}}\cg{\ell_1\lambda_1\ell_2\lambda_2|L\Lambda}
\cg{L\Lambda{S}\Sigma|I\mu} \,  \hat{L}\hat{S} \, \chi_{S\Sigma}
(1,2) \chi_{T\tau} (1,2)\nonumber\\
\times & [\pwt{1}{1}\pwt{2}{2}-(-)^{S+T}\pwt{1}{2}\pwt{2}{1}]\left\{
\begin{array}{ccc} \ell_1 & s & j_1 \\  \ell_2 & s & j_2 \\
L & S & I \\ \end{array} \right\}\ , \label{eqn:anti_wf_ls}
\end{align}
\end{widetext}
with $D_{\alpha}=1/\sqrt{2(1+\delta_{\beta_1\beta_2})}$ and where
the angular momentum and isospin couplings used are summarized in
Fig. \ref{fig:coupling}. The nucleon-wave functions $\pwt{}{i}$ are
\begin{align}
\pwt{}{i}=u_{\beta}(r_i)Y_{\ell\lambda}(\hr_i)\ . \label{eqn:wf_ls}
\end{align}

It is convenient to combine the statistical factors and $9j$
coefficient from the two-nucleon overlap with the appropriate
$jj$-coupled TNA to construct a set of $LS$-coupled TNA,
$\mathfrak{C}$, as
\begin{align}
\amp=& \,\hat{j}_1 \,\hat{j}_2 \, \hat{L} \, \hat{S} \,
\left\{ \begin{array}{ccc} \ell_1 & s & j_1 \\  \ell_2 & s & j_2 \\
L & S & I \\ \end{array} \right\}\,C_{\alpha}^{IT}\ ,\label{newTNA}
\end{align}
that satisfy the sum rule
\begin{align}
\sum_{LS}\left[\amp\right]^2=\left[C_{\alpha}^{IT}\right]^2\ .
\end{align}

Antisymmetry requires, for configurations $\alpha$ where the
nucleons originate from the same orbital, the $[n\ell j]^2$ cases,
that $L+S+T$ is odd. For nucleons originating from different
orbitals this is no longer the case; though for two nucleons from
spin-orbit partner orbitals, $[n\ell {j_<},n\ell {j_>}]$ with very
similar radial wave functions, the $L+S+T=$even amplitudes are also
expected to be significantly suppressed.

\subsection{Eikonal model of two-nucleon stripping}

As was developed previously \cite{TPB04,ToB06,STB09b}, we exploit
eikonal reaction dynamics. The elastic $S$-matrices describing the
absorptive interactions of the $A$-body core (in state $f$) and the
two nucleons with the target are calculated in the optical limit of
Glauber's multiple scattering theory \cite{Gla59,ATT96} assuming
that these projectile constituents travel on straight line paths in
the interaction field of the light target. The reaction is assumed
sudden, such that the projectile internal co-ordinates are frozen on
the timescale of this passing and interaction with the target. The
eikonal $S$-matrices are calculated from the nucleon- and heavy
residue-target interactions. These interactions were obtained by
double-folding the residue (core), nucleon ($\delta$-function) and
the target point particle densities with the usual effective
nucleon-nucleon interaction, as used elsewhere; e.g. \cite{TPB04}.

Following Refs. \cite{TPB04,ToB06,STB09b}, from the total absorption
cross section for the projectile-target system,
\begin{equation}
\sigma_{abs} = \frac{1}{\hat{J}_i^2} \sum_{M_i} \int{d\vec{b}}
\bra{\Psi_i} 1- |\s_f\s_1\s_2|^2 \ket{\Psi_i }\ ,
\end{equation}
that includes all events where one or more of the projectile
constituents are absorbed by the target, we can identify and extract
the two-nucleon stripping cross section terms,
\begin{equation}
\sigma_{str} = \frac{1}{\hat{J}_i^2} \sum_{M_i} \int{d\vec{b}}
\bra{\Psi_i} \ms{f}(1-\ms{1})(1-\ms{2}) \ket{\Psi_i }\ .
\label{eqn:2nstripping}
\end{equation}
As has been discussed elsewhere \cite{STB09b}, the two-nucleon
stripping probability $\mathcal{O}_{str}(b,b_1,b_2)=\ms{f}
(1-\ms{1}) (1-\ms{2})$ weights the impact parameters that contribute
to these stripping events. Stripping requires an absorptive
(inelastic) interaction of two nucleons with the target, but a
non-absorptive (elastic) or non-interaction of the heavy residue
with the target, and strongly localizes the reaction to grazing
collisions at the projectile surface. This simplifies our picture of
the reaction mechanism and of that part of the overlap function that
is probed in the knockout reaction. We also contrast the present
strong surface localisation of the reaction with the term {\em
peripheral} used by some authors to mean impact parameters that
sample only the extreme tail (the Whittaker or Hankel function
asymptotics) of the nucleon bound state wave function. This is
certainly not the case for those impact parameters selected by
$\mathcal{O}_{str}(b,b_1,b_2)$ at the projectile energies and with
the corresponding absorptive $S$-matrices of interest here.

Two further assumptions are made. The most important, but which is
supported by the stripping mechanism's selection of non-absorptive
or non-interactive events of the residue and target, is to assume
there is no dynamical excitation/change of state of the reaction
residue by $\s_f$ in the collision -- previously termed the spectator
core approximation. This being the case,
\begin{align}
\bra{\Phi^{(F')}(A)}& \ms{f} \ket{ \Phi^{(F)} (A)} = \ms{c}
\delta_{FF'}\ ,
\label{eqn:spectcore}
\end{align}
where the bra and ket integrate out the internal coordinates of the
residue and $\s_c$ is taken to be the residue ground state-target
elastic scattering $S$-matrix. A lesser assumption is the heavy-core
(no-recoil) approximation, that in the required integrals, such as
Eq. (\ref{eqn:2nstripping}), the residue impact parameter $b_c$,
entering $\s_f$ can be replaced by that of the center of mass of the
projectile $b$, i.e. $b_c \approx b$. For the stripping terms this
approximation is not in fact needed, as a change of integration
variable makes it unnecessary.

The result of these assumptions, together with the parentage
expansion for the two nucleon structure overlap, Eq.
(\ref{eqn:sm_overlap_ls}), is that the exclusive stripping cross
section to a given final state $f$ can be written
\begin{align}
\sigma_{str}^{(f)} = \int{d\vec{b}}\, \ms{c}  \frac{1}{\hat{J}_i^2}
\sum_{M_i M_f} & \bra{\Psi_i^{(F)} }(1-\ms{1}) \nonumber \\
& (1-\ms{2}) \ket {\Psi_i^{(F)} }\ . \label{final_2n}
\end{align}
The assumption that the nucleon $S$-matrices are spin-independent
allows one to carry out all spin coordinate sums, in preparation for
which we separate explicitly the nucleon position and spin variable
integrations, as
\begin{align}
\bra{\Psi_i^{(F)}}&\ldots\ket{\Psi_i^{(F)} } \nonumber\\
=& \int{d\vec{r}_1}\int{d\vec{r}_2}
\braket{\Psi_i^{(F)}|\ldots|\Psi_i^{(F)} }_{sp}\ ,
\end{align}
where the final bra-ket term denotes the spin integration.

Consideration of this (momentum-integrated) stripping cross section
using $LS$-coupling was made in Ref. \cite{TPB04}, there with an
emphasis on the reaction mechanism's lack of selectivity in the
total spin $S$ of the two nucleons. In our previous analysis of the
longitudinal momentum distributions \cite{STB09b} the only angular
momentum projections not able to be summed over algebraically were
those of the orbital angular momenta of the two nucleons.

This observation made, we now consider the residue longitudinal
momentum distributions in the $LS$-representation. The derivation
follows a similar pattern to that in the $jj$-coupled algebra and
begins from the $LS$-coupled, spin-integrated, modulus squared of
the two-nucleon overlap, averaged over initial projections $M_i$ and
summed over final projections $M_f$. Explicitly,
\begin{align}
\frac{1}{\hat{J}_i^2} \sum_{M_i M_f} &\braket{\Psi_i^{(F)}|\Psi_i^{(F)}}_{sp} =
\frac{1}{\hat{J}_i^2} \sum_{\substack{II'\mu\mu'\\\alpha\alpha'TT'}}C_{\alpha}^{IT}
C_{\alpha'}^{I'T'}\nonumber \\
\times & \sum_{M_i M_f }\cg{I\mu J_fM_f | J_i M_i}\, \cg{I'\mu'
J_fM_f | J_i M_i}  \nonumber\\ \times &\
\cg{T{\tau}T_f\tau_f|T_i\tau_i}\, \cg{T'{\tau}'T_f\tau_f|T_i\tau_i}
\nonumber\\
\times &\Braket{[\,\overline{\psi_{j_1'}(1)\otimes \psi_{j_2'}(2)}\,
]_{I'\mu'}^{T'\tau'} |
[\,\overline{\psi_{j_1}(1)\otimes\psi_{j_2}(2)}\,]_{I\mu}^{T\tau}
}_{sp}. \label{freddy}
\end{align}
On performing the sums over $M_i$ and $M_f$ the expression is
clearly incoherent in the coupled two-nucleon total angular momentum
$I \mu$, a consequence of the spectator-core approximation, Eq.
(\ref{eqn:spectcore}). Using the antisymmetric two-nucleon
$LS$-coupled forms of Eqs. (\ref{eqn:anti_wf_ls}) and
(\ref{newTNA}), and assuming the nucleon $S$-matrices are also
isospin-independent, we can perform the isospin sums with the result
that Eq. (\ref{freddy}) is also incoherent with respect to both $S$
and $T$. Finally, summing over the projections of $I$ and $S$ we
obtain the result, incoherent also in $L$ and $\Lambda$, namely
\begin{align}
\frac{1}{\hat{J}_i^2}&\sum_{M_i M_f}
\braket{\Psi_i^{(F)}|\Psi_i^{(F)}
}_{sp} \nonumber \\
= &\sum_{T}\cg{T{\tau}T_f\tau_f|T_i\tau_i}^2 \sum_{ILS\alpha\alpha'}
\frac{\amp\ampp D_{\alpha}D_{\alpha'}}{\hat{L}^2} \nonumber\\ &
\sum_{\Lambda \lambda_1\lambda_2\lambda_1'\lambda_2'}
\cg{\ell_1\lambda_1\ell_2\lambda_2|L\Lambda}
\cg{\ell_1'\lambda_1'\ell_2'\lambda_2'|L\Lambda} \nonumber\\
&[\pwpt{1}{1}\pwpt{2}{2}-(-)^{S+T}\pwpt{1}{2}\pwpt{2}{1}]^*\nonumber\\
&[\pwt{1}{1}\pwt{2}{2}-(-)^{S+T}\pwt{1}{2}\pwt{2}{1}]\ .
\label{eqn:spinav_overlap_ls}
\end{align}
The exclusive two-nucleon stripping cross section is then given by
use of this structure overlap information in  Eq. (\ref{final_2n}).
In the following we derive explicit expressions for the associated
exclusive momentum distributions in this $LS$-representation.

\subsection{Residue momentum distributions}

\begin{figure}[b]
\begin{center}
\includegraphics[width=0.9\columnwidth]{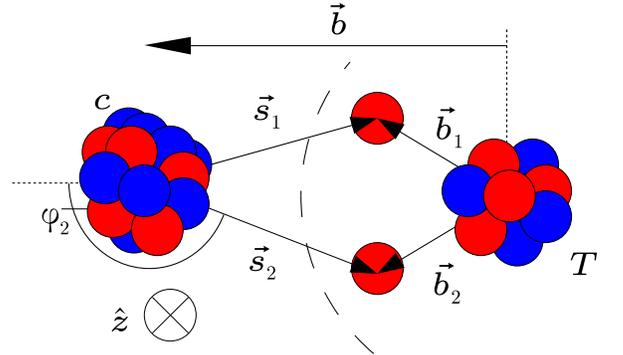}
\caption{\label{fig:coordinates} Schematic of the particle
coordinates used. Vectors $\vec{s}_i$ are the components, in the
plane perpendicular to the beam direction (the $z$-axis), of the
position vectors $\vec{r}_i$ of the knocked-out nucleons relative to
the core of nucleons to which they are initially bound. The two
nucleons have impact parameters $\vec{b_i} = \vec{b} + \vec{s}_i$
relative to the target nucleus.}
\end{center}
\end{figure}

Structurally, the expressions for the residue momentum distributions
in the $LS$-coupling scheme are similar to those using $jj$-coupling
\cite{STB09b}. The derivations also follow a largely parallel
procedure. The coordinate system used is reproduced in Fig.
\ref{fig:coordinates} for clarity of the following expressions.

The reaction samples the momentum content of the bound-state wave
functions of the stripped nucleons in the direction of the
projectile beam $\hat{\vec{k}}$ (i.e. the $z$-axis). For fixed
values of the $\vec{s}_i$, and hence fixed nucleon impact parameters
${b}_i= |\vec{b}+\vec{s}_i|,\ i=1,2$, this information is carried by
the functions
\begin{align}
\ray{}{i}&=\frac{\cfc{}}{\sqrt{2\pi}}\int^{+\infty}_{-\infty}\; dz_i
\; u_{\beta}(r_i) \; P_{\ell}^{|\lambda|}(\cos\theta_i)
\exp[i{\kappa_i}z_i]\ , \label{eqn:ray}
\end{align}
where $\kappa_i$ is the $z$-component of the momentum of nucleon $i$
in the projectile's rest-frame. We note that the notation for
$\ray{}{i}$ is changed from that of Ref. \cite{STB09b} consistent
with the notation used for the $u_{\beta}(r_i)$. The correct
weighting of the nucleon absorption probability with the azimuthal
angle $\varphi_i$, of $\vec{s}_i$, is carried by the functions $
\hay{}{}{i} \equiv \mathcal{H}_{\lambda \lambda'} (b,\vec{s}_i)$,
given by
\begin{align}
\hay{}{}{i}&=\int_0^{2\pi}d{\varphi_i}(1-|\s_i(|\vec{b}+
\vec{s}_i|)|^2)\exp[i\varphi_i (\lambda -\lambda')]\
,\label{eqn:hay}
\end{align}
with $\vec{s}_i$ the components of the $\vec{r}_i$ in the impact
parameter plane, i.e. $\vec{r}_i=\vec{s}_i+ z_i \hat{\vec{k}}$. The
remaining details of the derivation are completely analogous to
those in Ref. \cite{STB09b}, to which the reader is referred.

We obtain the projectile rest frame, stripping mechanism momentum
distribution as the incoherent $LS$ and isospin decomposition
\begin{align}
\frac{d\sigma_{str}^{(f)}}{d\kappa_c}  = &  \sum_{LST}
\frac{d\sigma_{LST}^{(f)}}{d\kappa_c} \nonumber \\
= &\sum_{T}\cg{T{\tau}T_f\tau_f|T_i\tau_i}^2
\sum_{LSI\alpha\alpha'}  \frac{2\amp\ampp
D_{\alpha}D_{\alpha'}}{\hat{L}^2}
\nonumber\\
& \int{d\kappa_1}\int{d\kappa_2}\;\delta(\kappa_c+\kappa_1+\kappa_2)
\int{d\vec{b}} \;|\s_c(b)|^2 \nonumber\\ & \sum_{\Lambda
\lambda_1\lambda_2\lambda_1'\lambda_2'}
\cg{\ell_1\lambda_1\ell_2\lambda_2|L\Lambda}
\cg{\ell_1'\lambda_1'\ell_2'\lambda_2'|L\Lambda} \nonumber \\
&\int{ds_1}s_1\int{ds_2}s_2\left[ direct - exchange\right],
\label{eqn:lssigma}
\end{align}
where the \emph{direct} term is
\begin{align}
direct =& \left\{ \hay{1}{1}{1} \ray{1}{1}
\rayp{1}{1} \right. \nonumber \\
\times & \left.  \hay{2}{2}{2} \ray{2}{2} \rayp{2}{2} \right\} ,
%\nonumber\\
%&+  \hay{1}{1}{2} \ray{1}{2}\rayp{1}{2} \nonumber \\
%&\hay{2}{2}{1} \ray{2}{1} \rayp{2}{1} \left.\rule{0pt}{12pt}\right\},
\end{align}
and the \emph{exchange} term,
\begin{align}
exchange=&  (-1)^{S+T} \left\{\rule{0pt}{12pt}
\hay{2}{1}{1}\ray{2}{1} \rayp{1}{1} \right. \nonumber \\
\times&\left. \hay{1}{2}{2}  \ray{1}{2} \rayp{2}{2} \right\} \ .
% \nonumber\\
%&+  \hay{2}{1}{2} \ray{2}{2} \rayp{1}{2}  \nonumber\\
%& \hay{1}{2}{1}  \ray{1}{1} \rayp{2}{1} \left. \rule{0pt}{12pt}\right\}\ .
\end{align}
It should be noted that Eq. (\ref{eqn:lssigma}) (that contains a
factor of 2) and these simplified forms for the \emph{direct} and
\emph{exchange} terms compared to \cite{STB09b}, assume that both
the integrals over the $\kappa_i$ and the $s_i$ will be carried out,
and so one is computing quantities that are completely symmetric in
the two nucleon coordinates. Also, unlike for the $jj$-coupled
scheme, no further re-coupling is required to reduce the angular
momentum algebra.

Physically, Eq. (\ref{eqn:lssigma}) shows that the sums of the
\emph{direct} and \emph{exchange} terms over the $\lambda$ are
independent of $I$. The resulting momentum distributions thus depend
explicitly on $L$ (and $S+T$ via the phase of the exchange term),
but not on $I$. We see that the significance of $I$ and the nucleon
total angular momenta $j_i$ is that they will determine the relative
strengths of the different $L$ and $S$ via the amplitudes $\amp$.
Thus it is $L$, and to a lesser extent $S$ and $T$, that will
determine the shape of the residue's momentum distribution. $I$ on
the other hand will be important in determining the relative
strengths of the $L$ and $S$ that contribute.

\section{Two-nucleon correlations \label{corr_sect}}

We observe that the expression for residue momentum distributions is
somewhat simpler when using $LS$-coupling, having a more transparent
angular momentum dependence. However, the dependence on the two
nucleon configurations, via $\alpha$ (and $\alpha'$), is still less
than transparent in Eq. (\ref{eqn:lssigma}). We attempt to elucidate
this important nuclear structure sensitivity by carrying out the
$\lambda$ projections sums. Before doing so we introduce and discuss
the two-nucleon joint position probability that summarizes both the
strength and the spatial localisation (and correlation) of the two
nucleons in the structure overlaps that affects the stripping yield.

\subsection{Two-nucleon joint position probability}

We consider the two-nucleon joint position probability relevant to
the removal reaction/transition to a given final state $f$, i.e.
\begin{align}
\rho_f(\vec{r}_1,\vec{r}_2) &= \frac{1}{\hat{J}_i^2}\sum_{M_i M_f}
\braket{\Psi_i^{(F)}|\Psi_i^{(F)}}_{sp}\ . \label{eqn:jpp}
\end{align}
While the production of a given residue final state by the
two-nucleon knockout mechanism will depend on the details of
$\rho_f(\vec{r}_1, \vec{r}_2$), specifically the extent to which
there is a spatial proximity of the two nucleons at the projectile
surface, its overall normalisation and the $LS$-composition of this
normalisation
\begin{align}
N_f =&
\int\!{d\vec{r}_1}\int\!{d\vec{r}_2}\;\rho_f(\vec{r}_1,\vec{r}_2)
\nonumber\\
=& \sum_{\alpha IT}
\left[\cg{T{\tau}T_f\tau_f|T_i\tau_i}C_{\alpha}^{IT}\right]^2\nonumber\\
=&\sum_{LS}\left\{\sum_{\alpha
IT}\left[\cg{T{\tau}T_f\tau_f|T_i\tau_i}\amp\right]^2\right\}=\sum_{LS}
N_f^{LS}, \label{norms}
\end{align}
are measures of the likely transition strength. In the case of a
single (dominant) two-nucleon structure configuration this $LS$
breakdown can also guide the relative strengths expected from the
different contributing $LS$ terms for a given final state. However,
when configurations are mixed or where the initial and final states
have different parity, interference effects may strongly affect
these relative strengths.

Since the projectile is assumed to traverse a straight line path in
the $z$-direction, it is useful for what follows, and also highly
intuitive, to construct the projection of the two-nucleon joint
position probability onto the impact parameter plane -- the plane
perpendicular to the beam direction -- by integration over the $z_i$
of the two nucleons,
\begin{align}
\mathcal{P}_f(\vec{s}_1,\vec{s}_2)=\int\!{dz_1}\int\!{dz_2}\;
\rho_f(\vec{r}_1,\vec{r}_2)\ . \label{eqn:projden}
\end{align}
The relevant spatial correlation for the reaction is now the degree
of localisation of the probability with respect to the two nucleon
coordinate projections $\vec{s}_i$ in this impact parameter plane.

In what follows the correlation of the two nucleons is concisely
expressed as a function of the angular separation, $\omega$, of
their position coordinates $\vec{r}_i$. Clearly, the $z_i$
integrated joint probability $\mathcal{P}_f (\vec{s}_1, \vec{s}_2)$
will see a smeared version of this correlation function since fixed
$\vec{s}_i$ will sample a range of $\omega$. However, since the
reaction is surface localized and the target is light (small) the
effective thickness in the $z_i$ will tend to be rather restricted
and $\mathcal{P}_f (\vec{s}_1, \vec{s}_2)$ will remain a useful
construct and intuitive link to the magnitudes of the two-nucleon
knockout cross sections.

As was indicated by Figs. 2 and 3 of Ref. \cite{senuf06}, and will
be emphasized here, the total angular momentum of the final state
and the detailed TNA of the wave function can strongly affect the
two-nucleon joint position probability, its projection and the
magnitude of the removal cross sections. The (shell-model)
structural correlations may also enhance or suppress particular
total orbital angular momenta and so may affect the residue momentum
distributions also.

\subsection{Angular Correlations}

Despite the relative simplifications introduced by $LS$-coupling,
the momentum distribution expression, Eq. (\ref{eqn:lssigma}),
remains a complicated weighted sum of wave function transforms.
Moreover, it still depends on the orbital angular momentum
projections. To clarify the underlying sensitivity to two-nucleon
correlations we simplify the spin-integrated two-nucleon
joint position probability of Eq. (\ref{eqn:jpp}) by summing out the
$\lambda$ projection labels.

The relevant terms we need to simplify are, for the direct terms of
Eq. (\ref{eqn:spinav_overlap_ls}),
\begin{align}
\Gamma_{\ell_1\ell_2\ell_1'\ell_2'}^{L}(\hr_1,\hr_2)=& \sum_{\Lambda
\lambda_1 \lambda_2\lambda_1'\lambda_2'}
\cg{\ell_1\lambda_1\ell_2\lambda_2|L\Lambda}
\cg{\ell_1'\lambda_1'\ell_2'\lambda_2'|L\Lambda} \nonumber\\
\times &Y_{\ell_1\lambda_1}(\hr_1) Y^*_{\ell_1'\lambda_1'}(\hr_1)
Y_{\ell_2\lambda_2}(\hr_2) Y^*_{\ell_2'\lambda_2'}(\hr_2).
\label{eqn:acf_begin}
\end{align}
Combining the spherical harmonics of the same argument, summing the
$\lambda$ projections, and using the spherical harmonics addition
theorem one obtains
\begin{align}
\acf =& (-1)^{L} \frac{\hl_1\hl_1'\hl_2\hl_2' \hat{L}^2}{(4\pi)^2}
\sum_{k} W(\ell_1\ell_2\ell_1'\ell_2';Lk) \nonumber \\ \times &
(-1)^k \cg{\ell_10\ell_1'0|k0} \cg{\ell_20\ell_2'0|k0}
P_k(\cos\omega), \label{eqn:acf_ls}
\end{align}
where $\omega$ is the angular separation of the two nucleons, i.e.
$\cos \omega =\vec{r}_1\cdot \vec{r}_2/r_1r_2$. A similar result can
be found in Ref. \cite{BBR67}.

The angular correlation function $\acf$ is seen to be independent of
the total angular momentum $I$ and of the individual angular momenta
$j_i$ of the nucleons. However, it depends explicitly on their
orbital angular momenta and on the total orbital angular momentum
$L$. The form written above is that for the {direct} terms of Eq.
(\ref{eqn:spinav_overlap_ls}). The {exchange} terms differ by a
phase due to the reordering of the angular momentum labels in the
exchange form of Eq. (\ref{eqn:acf_begin}), as is given below.

The radial behaviors associated with the direct and exchange terms
of the joint-probability density are
\begin{align}
U_{\alpha\alpha'}^{D}(r_1,r_2) = & \;u_{\beta_1}(r_1) \;
u_{\beta_2}(r_2) \; u_{\beta_1'}(r_1) \; u_{\beta_2'}(r_2)
\nonumber \\
+&\,  u_{\beta_2}(r_1) \; u_{\beta_1}(r_2) \; u_{\beta_2'}(r_1) \;
u_{\beta_1'}(r_2)\ , \nonumber\\ U_{\alpha\alpha'}^{E}(r_1,r_2) = &
\; u_{\beta_1}(r_1) \; u_{\beta_2}(r_2) \; u_{\beta_2'}(r_1) \;
u_{\beta_1'}(r_2) \nonumber
\\ +&\,  u_{\beta_2}(r_1) \; u_{\beta_1}(r_2) \; u_{\beta_1'}(r_1) \;
u_{\beta_2'}(r_2)\ .
\end{align}
In terms of these and the corresponding direct and exchange angular
correlation functions, the two-nucleon joint-probability density is
\begin{align}
\rho_f(\vecr_1,\vecr_2) = & \sum_{LST} \sum_{I\alpha\alpha'}
\frac{\amp\ampp D_{\alpha}D_{\alpha'}}{\hat{L}^2}
\cg{T{\tau}T_f\tau_f|T_i\tau_i}^2  \nonumber\\\times & \left[
U_{\alpha\alpha'}^{D}(r_1,r_2)\, \Gamma^{L,D}(\omega) \right.
\nonumber \\ -& \left. (-)^{S+T} U_{\alpha\alpha'}^{E}(r_1,r_2)\,
\Gamma^{L,E}(\omega) \right], \label{eqn:final_density}
\end{align}
with $\Gamma^{L,D}(\omega)\equiv\acf$ given by Eq. (\ref{eqn:acf_ls}) and
\begin{equation}
\Gamma^{L,E}(\omega) = (-)^{\ell_1'+\ell_2'-L}\, \acfx\,.
\end{equation}

It is clear therefore that the angular correlation function dictates
how the spatial correlations change with angular momentum coupling,
and that $L$ is crucial, the $U(r_1,r_2)$ being dependent on the
$\beta_i$ but independent of the angular momentum coupling. Clear
also is that, in cases where the radial wave functions for all
active orbits are similar, the angular correlation function alone
will determine the differences in residue momentum distributions for
the different possible angular momentum couplings. As was discussed
earlier, these differences, generated at the angular correlation
function and the two-nucleon density level, will be more distinct
than in the projected density, Eq. (\ref{eqn:projden}) where fixed
co-ordinate pairs $(\vec{s}_1,\vec{s}_2)$ sample a range of angular
separations $\omega$ and so will smear the spatial correlations.

Uncorrelated two nucleon models, discussed in the Introduction and
Refs. \cite{BBC03,TPB04}, that neglect antisymmetrization, angular
momentum coupling and parentage coefficients lead to a constant,
$\omega$-independent correlation function. For two nucleon removal
from a single $[s_{1/2}]^2$ configuration, the angular correlation
function is also seen to be $\omega$-independent ($k=0$) and the
uncorrelated (see Ref. \cite{STB09b}) and fully correlated residue
momentum distributions will be identical.

\subsection{Cross shell excitations}

Here we consider briefly the implications for two-nucleon knockout
from configurations with $\beta_1$ and $\beta_2$ of different
parity. It is well established that the addition of shell-model
configurations with $1\hbar\omega$, $3\hbar\omega$ single particle
excitations are required to obtain a high degree of surface pairing
(see e.g. \cite{Pin84,JaL83,CIM84a,CIM84b,IJM89,TTD98}).

We obtain a similar result here, by considering the symmetry of the
angular correlation function about $\omega=\pi/2$. In $\acf$, only
the Legendre polynomial depends on $\omega$, with the property that
$P_k(\cos[\pi-\omega])= (-1)^kP_k(\cos\omega)$. Since the values of
$k$ are restricted to be odd or even by the parity Clebsch Gordan
coefficients, the angular correlation will be even about $\pi/2$ for
$\pi_\ell \pi_{\ell'}=+1$ and odd about $\pi/2$ for $\pi_\ell
\pi_{\ell'}=-1$. In the absence of single-particle excitations of
the kind $1\hbar\omega$, the probability for finding the nucleon
pair with angular separation $\omega=0$ and $\pi$ are equal and a
high degree of two-nucleon pair/cluster structure will not be
obtained.

So, pair correlations will be enhanced in cases when there is mixing
between two-nucleon configurations where the orbital angular momenta
are of different parity.  Whether the interference is constructive
or destructive will depend on the sign of $\acf$ near $\omega=0$,
the relative signs of the $9j$ coefficients, and the relative signs
of the TNA. A specific two configuration example will be presented
in the Section \ref{sec:ti54} below.

These results are quite general in that they do not depend on the
pair total angular momentum $I$; enhancements in the spatial
correlations in the two-nucleon density may be found for $I \ne 0$.

\section{Illustrative examples \label{examp_sect}}

Previous calculations of exclusive two-nucleon removal residue
momentum distributions noted a strong sensitivity to the total
angular momentum of the removed nucleon pair. Here, by writing this
momentum-differential cross section in $LS$-coupling, and by a
consideration of the angular correlations inherent in the
two-nucleon joint probability function, it becomes apparent that the
crucial sensitivity of this observable is to the total orbital
angular momentum values, $L$, contributing to the transition. These
different $L$ components will contribute incoherently to the cross
section yields and their momentum distributions. These theoretical
observations and the resulting sensitivity of the momentum
distribution observable offers the potential to probe more subtle
features of the nucleon pair's configurations and the correlations
present in the shell-model wave functions used.

A generic first example will arise if the predominant two-nucleon
configuration populating a given final state involves one of the
nucleons in an $s$-wave orbital. In this case the total orbital
angular momentum is restricted to the orbital angular momentum of
the second active orbit, $L=\ell$, and thus $L$ is pure. It is
expected therefore that there can be distinct differences in the
momentum distributions, even for states of the same total angular
momentum $I$. For example, for two $3^+$ final states built from
$[g_{7/2}, s_{1/2}]$ and from $[d_{5/2},s_{1/2}]$. 
More generally, even where there is significant
mixing and several active configurations, the structure of specific
states in the spectrum can be rather $L$-pure. So, the reaction will
proceed by a particular $L$ with a momentum distribution that is
characteristic of this structure.

In the following we discuss specific examples from different $A$ and
$Z$ regions of the nuclear chart. In each example the nucleon bound
state radial wave functions required for the two-nucleon overlaps
are calculated using a Woods-Saxon potential well with a spin-orbit
term of depth 6 MeV and a diffuseness parameter $a_0$=0.7 fm. Unless
stated otherwise, the geometries (the radius parameters $r_0$) of
the potential wells in each case were adjusted to reproduce the root
mean square radii and the separation energies of spherical
Hartree-Fock calculations using the Skyrme (SkX) interaction
parameterization \cite{brown98} for the active orbitals in question.
The specific procedure was detailed in Ref. \cite{Gad08}. These
fitted geometries are then used to calculate the radial wave
functions needed using the empirical, effective nucleon separation
energies. Where required, shell-model calculations are performed
using the code {\sc oxbash} \cite{BEG04}. The model spaces and
interactions used are specified for each case studied, below.

\subsection{p-shell example: \nuc{12}{C}(-np)}

Here we consider the removal of a ($T=0, 1$) neutron and proton
($np$) pair from \nuc{12}{C} at 2100 MeV/nucleon on a \nuc{12}{C}
target.   The proton
and neutron orbits are taken to be identical with radial wave
functions calculated in a Woods-Saxon potential, using an average
nucleon charge $\bar{Z}=0.5$.  The geometry of the Woods-Saxon
potential was fixed with $r_0=1.31$ fm, $a_0=0.7$ fm. 
Both the \nuc{10}{B} residue and \nuc{12}{C} target were assumed 
have Gaussian shaped mass distributions, with rms radii 2.30
and 2.32 fm respectively.
The isospin format TNA are calculated
using {\sc oxbash} in a $p$-shell model space using the {\sc wbp}
interaction \cite{WaB92}, as in previous studies \cite{BHS02,TBP04}.
A more complete consideration of two-nucleon removal from \nuc{12}{C} will
be discussed in a forthcoming paper \cite{ToS10}.

As a specific example, we consider the first and second $T=0$,
\nuc{10}{B}($1^+$) final states. The TNA for these states are shown
in Table \ref{tbl:b10_tna}. The relative magnitudes of the
contributing two-nucleon configurations to these states are
different and it is of interest to consider how these differences
might affect the cross sections and their momentum distributions.
The sum of the squared TNA for the first and second states are 1.45
and 1.47, respectively, thus in the absence of interference terms
the incoherent sum of contributions from each of these
configurations would yield very similar cross sections.

\begin{figure}[tb]
\includegraphics[angle=-90,width=\columnwidth]{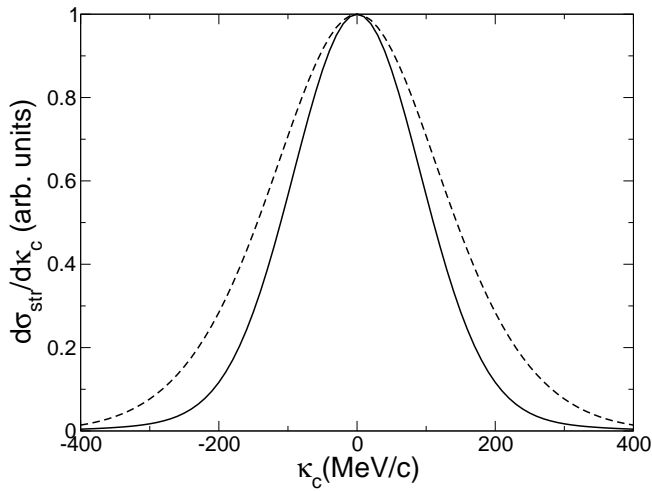}
\caption{Normalized residue momentum distributions for the first
(solid) and second (dashed) \nuc{10}{B}($J_f$=1$^+$) states
populated in $np$ knockout from \nuc{12}{C} at 2100 MeV per nucleon.
Though the same two-nucleon configurations contribute to each state,
the differently weighted TNA result in distinct momentum
distributions. \label{fig:one_plus_dkc}}
\end{figure}

That this is not the case is shown by the calculated two-nucleon
stripping cross sections presented in Table \ref{tbl:sigma_LS}.  The
calculated momentum distributions are also rather different, as is
shown in Fig. \ref{fig:one_plus_dkc}.

\begin{table}[tbp]
\caption{Isospin format two nucleon amplitudes for the first and
second \nuc{10}{B}($1^+$) $T=0$ states populated by neutron-proton
removal from \nuc{12}{C}.  The relative strengths of the two
nucleon amplitudes are different for the two states.
\label{tbl:b10_tna}}
\begin{ruledtabular}
\begin{tabular}{cccc}
$J_f^\pi$ & $[1p_{3/2}]^2$ & $[1p_{1/2},1p_{3/2}]$
& $[1p_{1/2}]^2$
\\
\hline
1$_1^+$   & ~0.69899 & ~0.97868 & $-$0.01067 \\
1$_2^+$   & $-$1.13385 & ~0.22886 & ~0.36314 \\
\end{tabular}
\end{ruledtabular}
\end{table}

\begin{table}[tbp]
\caption{Two-neutron stripping cross sections $\sigma_{LS}$ for
populating the first two $T=0, J_f= 1^+$ final states in
$\nuc{10}{B}$ for a \nuc{12}{C} beam energy of 2100 MeV per
nucleon. All cross sections are in mb.\label{tbl:sigma_LS}}
\begin{ruledtabular}
\begin{tabular}{cccccc}
$J_f^\pi$ & $\sigma_{01}$  & $\sigma_{10}$ & $\sigma_{11}$ &
$\sigma_{21}$ & $\sigma_{str}$ \\ \hline
1$_1^+$  & 2.41 & 0.00 & 0.00 & 0.06 & 2.47  \\
1$_2^+$  & 0.60 & 0.59 & 0.00 & 0.63 & 1.81  \\
\end{tabular}
\end{ruledtabular}
\end{table}

These differences can be understood by reference to the projected
two-particle joint position probabilities for the two states, which are strikingly
different. The first $1^+$ state shows strong spatial localisation
of the two-nucleons, favorable for the two-nucleon removal cross
section. Both example position probabilities manifest the expected symmetry about
a nucleon angular separation of $\phi_{12}=\pi/2$, since the model
space is restricted to the $p$-shell and the active orbitals have
the same parity.

\begin{figure}[tb]
\includegraphics[width=0.9\columnwidth]{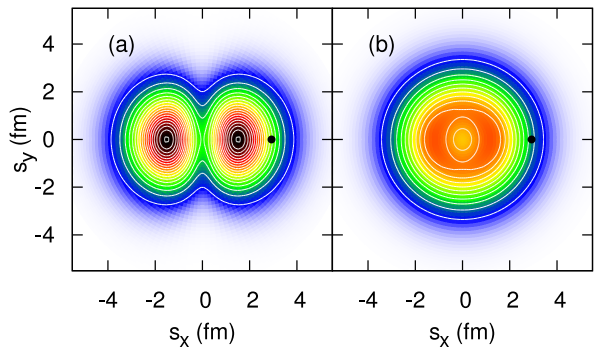}
\caption{(Color online) Impact parameter plane-projected joint position probabilities for (a) the
first and (b) second $T=0$ \nuc{10}{B}(1$^+$) states populated
via $np$ knockout from \nuc{12}{C}. The plot shows the impact
parameter plane probability distribution of nucleon 2 for nucleon 1
positioned at $s_x=2.9$ fm, $s_y=0$ fm.  The spatial correlations
of the nucleon pairs in these two states are fundamentally
different, leading to markedly different momentum distributions; see
Fig. \protect\ref{fig:one_plus_dkc}.  The color scale
(white-blue-green-yellow-red-black) is common to both plots.
\label{fig:one_plus_projden}}
\end{figure}

We can extend this $p$-shell example further to illustrate the
potential for large sensitivity to the underlying structure.  It is
clear from Eq. (\ref{eqn:final_density}) that within a $p$-shell
model space the relative strengths of different $LS$ combinations
are determined solely by the TNA and the nucleon configurations
involved. So, neglecting any minor differences in the $p$-wave
radial wave functions, due to spin-orbit splitting, the entire
square bracketed term in Eq. (\ref{eqn:final_density}) is
independent of the total angular momenta $\{j_i\}$ and, in the
present model space, independent of the configurations $(\alpha,
\alpha')$ of the pair. It follows that the weight of each $L$ and
$S$ term in a state of given $I$ and $T$ is proportional to
\begin{align}
P^{IT}_{LS}= \sum_{\alpha\alpha'} \amp \ampp D_{\alpha}
D_{\alpha'}\,.
\end{align}
In the sprit of studying the extremes of possible sensitivity of the
momentum distributions, we may force any one of these $P^{I T}_{L
S}$ to be zero, and solve for the relative strengths and phases of
the $\amp$ and $C^{IT}_\alpha$ needed to achieve this.

\begin{figure}[tb]
\includegraphics[angle=-90,width=\columnwidth]{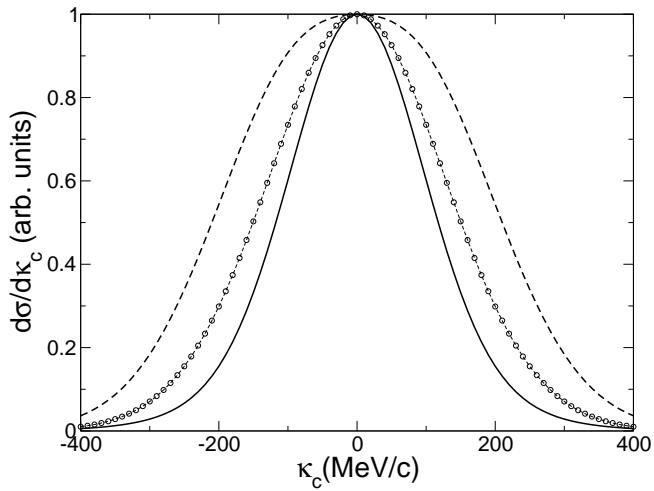}
\caption{Theoretical \nuc{10}{B} residue momentum distributions at
2100 MeV per nucleon. Shown are the expectations for two $I=1$ $T=0$
states where the TNA have been tailored to exclude $L=2, S=1$ (solid
line) and $L=0, S=1$ (dashed line) contributions from the
two-nucleon wave function, respectively. The line with open circles
shows the expectation for an $I=2$ $T=1$ final state where the TNA
were similarly chosen to eliminate $L=2, S=1$ contributions.
\label{fig:c12ls}}
\end{figure}

Fig. \ref{fig:c12ls} illustrates such examples for assumed $I=1$,
$T=0$ states populated via the configurations $\alpha_1=[1p_{1/2}]^2$
and $\alpha_2=[1p_{3/2},1p_{1/2}]$. Calculations for two sets of TNA
are shown; one set chosen to eliminate $L=0, S=1$ contributions
(requiring $C_{\alpha_1}^{10}=-4C_{\alpha_2}^{10}$, dashed curve)
and the other to eliminate $L=2, S=1$ contributions (requiring
$C_{\alpha_1}^{10} =C_{\alpha_2}^{10}/2$, solid curve). These
different $1^+$ state TNA produce wide and narrow residue momentum
distributions, respectively, the difference in the FWHM widths being
almost a factor of two. The figure also shows the $I=2, T=1$
momentum distribution (open circles), populated via $\alpha_2
=[1p_{3/2},1p_{1/2}]$ and $\alpha_3=[1p_{3/2}]^2$, where the TNA were
chosen to eliminate $L=2, S=1$ contributions (requiring
$C_{\alpha_3}^{21} = -\sqrt{2} C_{\alpha_2}^{21}$). Once again this
gives a relatively narrow distribution and, moreover, this $I=2$
distribution is \emph{narrower} than that for the ($L=0$ excluded)
$I=1$ distribution (dashed curve) described above. These examples
break the tie between the width of the momentum distribution and the
$I$ value of the transferred pair. Whether or not nuclear states
with these TNA are physically realized, these limiting cases
demonstrate how details of the microscopic structure of a given
state may strongly influence the shapes and widths of the expected
residue momentum distributions.

Similarly, we note the expectation that transitions to $I=2, T=0$
and $I=3, T=0$ states of \nuc{10}{B} will, within a $p$-shell model
space, yield identical theoretical momentum distributions since both
transitions are pure $L=2$ in nature. In this instance also the
width of the momentum distribution does not provide a direct measure
of $J_f=I$.

We note that consideration has only been given to the direct
population of \nuc{10}{B}. In principle, indirect population by
single nucleon knockout followed by evaporation of the unlike
nucleon may be possible, although we expect this indirect, two-step
pathway to be very weak, due to the large nucleon separation
energies in the relevant $A=11$ systems and the very small predicted
shell model strength for one nucleon removal to states above these
first $A=11$ nucleon thresholds. High precision (stable beam)
observations of final state exclusive \nuc{10}{B} momentum
distributions would clarify such aspects of the reaction mechanism
that are currently assumed.

\subsection{sd-shell: \nuc{28}{Mg}(-2p) and \nuc{26}{Si}(-2n)}

Exotic nuclei with valence nucleons in the $sd$-shell have been the
focus of several two-nucleon removal experiments, studying the
evolution of structure away from the valley of $\beta$-stability.
The initial and final state structures are often well described
within conventional $sd$-shell model space calculations offering
good test cases for studies of the reaction mechanism. Details of
their residue momentum distributions could offer an additional test
of the shell model and the reaction mechanism in this region.

We first review the two-proton knockout from \nuc{28}{Mg} at 83.2
MeV/nucleon on a \nuc{9}{Be} target, previously studied in Refs.
\cite{BBC03,ToB06,STB09b}. To date, this is the only experimental
example with measured final-state exclusive \nuc{26}{Ne} momentum
distributions. Four states were populated, being the $0^+$ ground
state, the first and second $2^+$ states and the first $4^+$ state.
Previous work demonstrated the significant difference between the
ground state and $4^+$ residue momentum distributions, despite
strong experimental (reaction target) broadening of the measured
distributions.

We comment here on the effects on the 2.02 and 3.70 MeV $2^+$ state
momentum distributions of the subtle differences in their TNA,
tabulated in Ref. \cite{TPB04}. To remove the small difference in
the average separation energies of the protons for the two states,
calculations used identical radial wave functions, but this binding
effect is in practice negligible. The calculated widths of the
residue momentum distributions are different by $\sim10\%$. Clearly
a higher statistics experiment would be required to examine this
difference predicted by the shell model. There are however other
examples where the $sd$-shell model predicts TNA that exhibit a
larger degree of sensitivity, as e.g. the following.

A second specific example is the two-neutron ($T=1$) knockout from
\nuc{26}{Si}, measurements for which were reported in Ref.
\cite{YOG06}; made at 109 MeV per nucleon on a \nuc{9}{Be} target.
Details of the nucleon radial wave functions and $S$-matrices can
be found in Ref. \cite{ToB06}.
Populations of two excited states in \nuc{24}{Si} were observed, the
first $2^+$ state at 1.86 MeV, and a state at 3.41 MeV corresponding
to a theoretically-predicted $(2^+,4^+)$ doublet, with theoretical
excitation energies of 3.867 and 3.962 MeV. The cross sections for
these measured and theoretical states were analyzed \cite{ToB06}
assuming that the second excited state was the second $2^+$ state.
Momentum distributions, if available, would easily distinguish
between such $I=2$ and $I=4$ possibilities. Our interest here is
more subtle. We consider the expected differences in the momentum
distributions of the two $2^+$ states arising from their underlying
$sd$-shell model structures.
\begin{table*}[tbp]
\caption{USD shell-model TNA for the first two $2^+$ states in
\nuc{24}{Si}, populated in two-neutron knockout from \nuc{26}{Si}.
\label{tbl:si26_tna}}
\begin{ruledtabular}
\begin{tabular}{cccccc}
$J_f^\pi$ & $[1d_{5/2}]^2$ & $[1d_{5/2},1d_{3/2}]$ & $[1d_{3/2}]^2$ &
$[2s_{1/2},1d_{3/2}]$ & $[2s_{1/2}, 1d_{5/2}]$  \\ \hline
2$_1^+$  & $-$0.70074 & ~0.43499 & ~0.00594 & $-$0.00188 & $-$0.02781 \\
2$_2^+$   & $-$0.38021 & $-$0.12354  & $-$0.12945 & $-$0.15876 & $-$0.58292\\
\end{tabular}
\end{ruledtabular}
\end{table*}

The TNA were calculated using {\sc oxbash} within an $sd$-shell
model space using the USD interaction \cite{Wil84} and are presented
in Table \ref{tbl:si26_tna}. The TNA calculated using the USDA and
USDB interactions \cite{BrR06} were found to be very similar to the
USD values. Both states have mixed $sd$-shell configurations.
Inspection of the TNA might suggest that since the second state has
a stronger $[2s_{1/2}, 1d_{5/2}]$ configuration it may favor
$L=2$ more strongly, but there is significant mixing.

\begin{figure}[tb]
\includegraphics[angle=-90,width=\columnwidth]{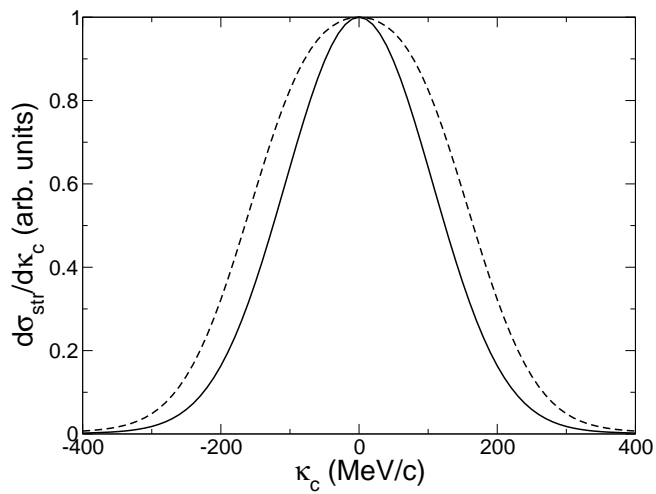}
\caption{Normalized residue momentum distributions for the first
(solid) and second (dashed) \nuc{24}{Si}(2$^+$) states populated in
$2n$ knockout from \nuc{26}{Si}. Though the same $sd$-shell
two-nucleon configurations contribute to each state, their TNA
result in distinct $L$ makeup and momentum distributions.  The
(FWHM) peak widths are are 250 and 330 MeV/c, respectively.
\label{fig:si26_2+}}
\end{figure}

Despite the strong mixing in both states, the shell-model TNA
predict each state to be populated predominately by a single and
distinct total orbital angular momentum $L$, $L=1$ and $L=2$,
respectively. The calculated $LS$-coupled two-nucleon stripping
partial cross sections reveal this, as are shown in Table
\ref{tbl:si26_sigmaLS2}.
\begin{table}[tbp]
\caption{Two-neutron stripping cross sections, $\sigma_{LS}$, for
population of the first two $2^+$ final states in $\nuc{24}{Si}$.
All cross sections are in mb. \label{tbl:si26_sigmaLS2}}
\begin{ruledtabular}
\begin{tabular}{cccccc}
$J_f^\pi$ & $\sigma_{11}$  & $\sigma_{20}$  & $\sigma_{21}$ &
$\sigma_{31}$ & $\sigma_{str}$
\\
\hline
2$_1^+$  &  0.17  &  0.02  &  0.00 &  0.00 &  0.19\\
2$_2^+$   &  0.01 & 0.17 & 0.01 & 0.00 & 0.19\\
\end{tabular}
\end{ruledtabular}
\end{table}
The dominance of $L=1$ and $L=2$ in these states generates the
significantly different $2^+$ state momentum distributions of Fig.
\ref{fig:si26_2+}, the $2_2^+$ state having a 30\% larger width.
Exclusive measurements for these states would not only clarify if
the second excited state is the $2_2^+$, but could also confirm the
$L=2$ dominance prediction of the $sd$ shell-model calculations.

To consolidate our understanding of such sensitivity, we consider a
further simplified example where a single configuration is expected
to dominate. We consider the two configurations $[1d_{5/2}]^2$ and
$[1d_{5/2}, 1d_{3/2}]$, both of which can contribute to 4$^+$
states. States with such simple configurations may not be realized
in \nuc{24}{Si}, since 4$^+$ states in \nuc{24}{Si} are thought to
be unbound, but the example will serve to illustrate the expected
differences that may occur elsewhere in the $sd$-shell.

We construct the TNA as $C_{\alpha}^{41}=\sqrt{5/3}$, such that
$N_f=1$, see Eq. (\ref{norms}), and the resulting $LS$-decomposition
of strengths is given in Table \ref{tbl:si26_sigmaLS4}. It is very
clear that the $[1d_{5/2}, 1d_{3/2}]$ configuration weights
$L=4$ significantly more strongly than does $[1d_{5/2}]^2$ and the
expectation is a wider momentum distribution. As noted in Section
\ref{sec:overlap}, in this case we would expect $L+S+T$=even
contributions to be significantly suppressed due to the two-neutron
antisymmetry and the similarity of the radial wave functions for the
active spin-orbit partner orbitals. This is indeed the case, as
demonstrated by the stripping cross sections of Table
\ref{tbl:si26_sigmaLS4}. The estimated strengths, $N_f^{LS}$, are
seen to give a reasonable guide to the expected cross sections
for these single configuration examples.

\begin{table}[tbp]
\caption{$LS$-coupled strengths, $N_f^{LS}$, and $LS$ partial
stripping cross sections for pure $[1d_{5/2}]^2$ and pure $[1d_{5/2}, 1d_{3/2}]$ 
two-neutron configurations populating
$4^+$ final states in $\nuc{24}{Si}$. The cross section for $L=4$,
$S=1$ is negligible, see text. \label{tbl:si26_sigmaLS4}}
\begin{ruledtabular}
\begin{tabular}{cccccccc}
$\alpha$ & $N_f^{31}$  & $N_f^{40}$  & $N_f^{41}$ & ~ &
$\sigma_{31}$ & $\sigma_{40}$ & $\sigma_{41}$
\\
\hline
$[1d_{5/2}]^2$                &  0.8 & 0.2 & 0.0 & ~ & 0.23 & 0.09 & 0.00 \\
$[1d_{5/2}][1d_{3/2}]$   &  0.1 & 0.4 & 0.5 & ~ & 0.06 & 0.35 & 0.00 \\
\end{tabular}
\end{ruledtabular}
\end{table}

\begin{figure}[tb]
\includegraphics[angle=-90,width=\columnwidth]{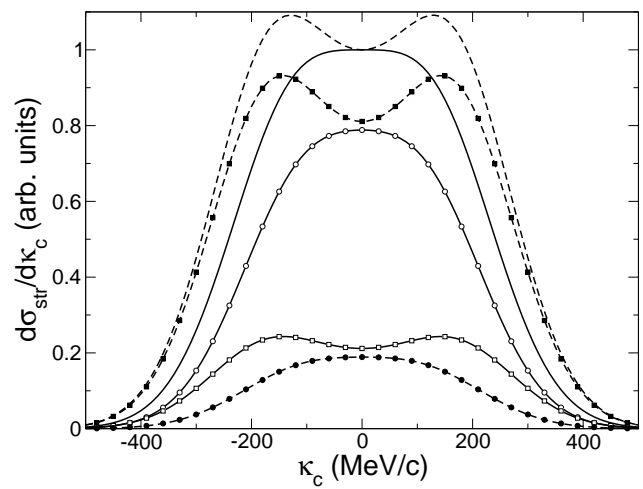}
\caption{Examples of the projectile rest frame residue momentum
distributions for $I=4$ states in the \nuc{26}{Si}($-2n$) reaction,
arising from different neutron pair configurations. The solid lines
and open points assume $[1d_{5/2}]^2$ neutron removal, the dashed
lines and solid points assume $[1d_{5/2},1d_{3/2}]$ neutron
removal. The full distributions are normalized to 1 at $\kappa_c=0$,
with each contributing $L$ partial distribution scaled by the same
factor. Circles show $L=3$, $S=1$ contributions and squares show
$L=4$, $S=0$, with the total shown by the line. The different
relative strengths of $L=3$ and $L=4$ for the two configurations
generates significantly different \nuc{24}{Si} momentum
distributions for states of the same $I$ (the dashed and solid
curves). \label{fig:si26_4+}}
\end{figure}

The results of the calculations, shown in Fig. \ref{fig:si26_4+},
confirm the differences in the momentum distributions expected from
our simple consideration of the $N_f^{LS}$. Again, the specifics of
the underlying structure predict considerable and observable
differences in the expected residue momentum distributions.

\subsection{Cross shell: \nuc{54}{Ti}(-2p)\label{sec:ti54}}

This ($T=1$) reaction, reported in Ref. \cite{GJB06}, demonstrated
the potential for two-nucleon knockout to probe cross-shell proton
excitations in neutron rich nuclei. In particular, a
\nuc{52}{Ca}(3$^-$,\,3.9 MeV) state was populated in two proton
removal from \nuc{54}{Ti}(0$^+$) on a \nuc{9}{Be} target at 72 MeV
per nucleon.
Details of the eikonal $S$-matrices and nucleon radial
wave functions can be found in Ref. \cite{GJB06}.  
Previous theoretical estimates for the 3$^-$ state
yield assumed pure $[1f_{7/2}, 2s_{1/2}]$ or $[1f_{7/2}, 1d_{3/2}]$
configurations, providing an estimated upper limit for the cross
section to this state as an incoherent sum of these contributions.

\begin{figure}[t]
\includegraphics[width=\columnwidth]{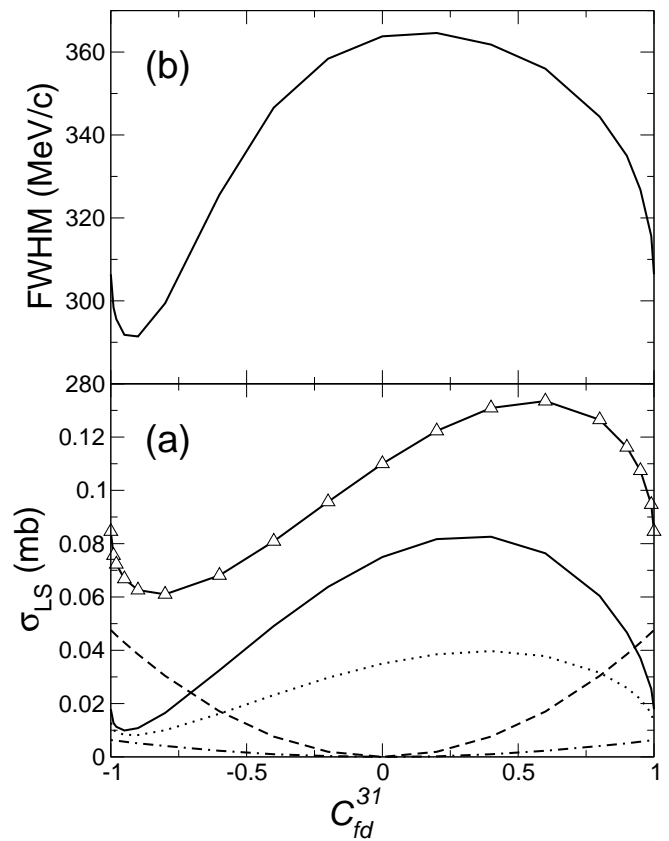}
\caption{(a) Two-nucleon stripping cross sections and (b) the
full-width at half-maximum of the residue momentum distribution 
when populating the \nuc{52}{Ca}(3$^-$, 3.9 MeV) state. These
observables are shown as a function of the amplitude $C_{fd}^{31}$.
The (positive) amplitude $C_{fs}^{31}$ of the second configuration
is given by Eq. \ref{eqn:fsamp}.  The lower panel shows the partial
cross sections with \{$L,S$\}, for the values \{$2,1$\} (dashed),
\{$3,0$\} (solid), \{$3,1$\} (dotted) and \{$4,1$\} (dot-dashed).
The total stripping cross section is shown by the solid line with
open triangles). \label{fig:ti54mix}}
\end{figure}

Taking instead a coherent sum will give (a) a different total cross
section, and (b) a different residue momentum distribution. Here we
assess the expected sensitivity to the relative strengths and phases
of these two configurations. We calculate the two-proton stripping
cross sections and momentum distributions as a function of the TNA
for $[ 1f_{7/2}, 1d_{3/2}]$ removal, $C_{fd}^{31}$, and for 
$[1f_{7/2}, 2s_{1/2}]$ removal, $C_{fs}^{31}$. For either of these
pure configurations the stripping cross sections scale with $[C_{f
\ell}^{IT}]^2$.  To maintain an overall scaling when the
configurations are mixed, the two amplitudes are adjusted such that
\begin{align}
[C_{fs}^{31}]^2+[C_{fd}^{31}]^2 =1\ , \label{eqn:fsamp}
\end{align}
with $C_{fs}^{31}$ assumed positive. The total incoherent strength
thus remains constant. We calculate the stripping cross sections for
each contributing $LS$ combination and the full-width half-maximum
(FWHM) for the residue momentum distributions. This is not the whole
story for the momentum distribution -- there are also subtle changes
of shape beyond the nominal width -- but this FWHM width provides a
guide to the expected behavior. The resulting calculations are shown
in Fig. \ref{fig:ti54mix}. A few points follow immediately; the $[
1f_{7/2}, 2s_{1/2}]$ configuration only contributes to the $L=3$
cross section, giving no interference with $L=2$ and $L=4$. So,
these latter terms are simply proportional to $[C_{fd}^{31}]^2$ and
are zero at the centre of the plot. The $L=4$ contributions are also
generally weak and the overall width of the residue momentum
distribution is largely determined by the relative strengths of the
$L=2$ and $L=3$ contributions.

Both the cross section and FWHM of the momentum distribution show a
strong sensitivity to the mixing of the two configurations; the
cross section varies by a factor of two and the width of the
momentum distribution by 25\%. It is clear that the underlying
structure and the relative strengths of the two-nucleon amplitudes
are critical to determining both the removal cross section and the
shape of the momentum distribution.
Of interest are the extremes of the plot, with $|C_{fd}^{31}|
\approx 1$. Here both the two-nucleon removal cross sections and
momentum distribution widths are acutely sensitive to the small
admixtures of the $[ 1f_{7/2}, 2s_{1/2}]$ configuration, but also
strongly dependent on its sign.  If the two amplitudes are of
opposite phase then both the cross section and width decrease
rapidly. Conversely, they increase rapidly if in phase. This is
indicative of a sensitivity to small cross-shell admixtures in many
cases.

\begin{figure}[t]
\includegraphics[width=0.9\columnwidth]{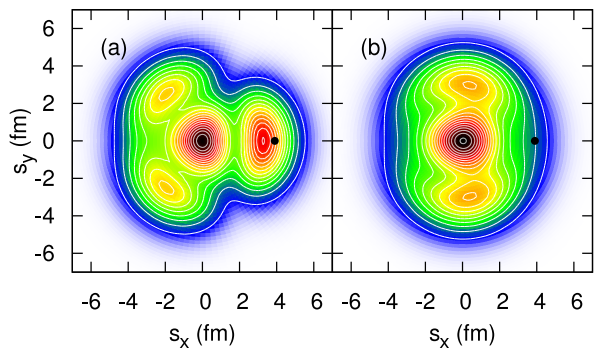}
\caption{(Color online) Impact parameter plane-projected two-nucleon joint position
probabilities, for $L=3$, $S=0$, for (a) $C_{fd}^{31}=\sqrt{2}/2$  and (b) $C_{fd}^{31}=-\sqrt{2}/2$
for the
\nuc{52}{Ca}(3$^-$, 3.9 MeV) state populated in two-proton knockout
form \nuc{54}{Ti}. The plot shows the probability distribution of
nucleon 2 when nucleon 1 is positioned at the back circle.
The color scale (white-blue-green-yellow-red-black) is the same
for both plots.
The source of the differences in the calculated cross sections (see Fig.
\ref{fig:ti54mix}) for these two choices of TNA is evident in the
pair correlations manifest in these projected two-particle
joint position probabilities  Note the asymmetry at $\varphi_{12}=90^\circ$
(i.e. $s_x=0$), induced by mixing of different parity orbitals.
\label{fig:ti54jpp}}
\end{figure}

We now contrast the impact parameter plane-projected two-nucleon
joint probability distributions, for values of $C_{fd}^{31}=\pm
\sqrt{2}/2$, in Fig. \ref{fig:ti54jpp}. The difference in this
cross-shell case is now striking and it is clear that taking the
amplitudes to be in phase (the +ve choice) enhances the two-nucleon
spatial correlations, that then drives the significantly larger
two-nucleon knockout cross section that is calculated and differences
in the residue momentum distribution.

As is clear from Fig. \ref{fig:ti54mix}, a precise measurement of
the residue momentum distribution for this reaction would allow an
estimate of the relative strengths and phases of the amplitudes of
the two assumed active two-nucleon (cross-shell) configurations. The
knockout cross sections themselves are also shown to depend strongly
on the mixing. To date, analyses of two-nucleon knockout from exotic
(asymmetric) systems have shown that the theoretical cross sections
overestimate those measured experimentally by of a factor of about
two, quantified as $R_s(2N )=\sigma_{expt} / \sigma_{theor} \approx
0.55$; see e.g. Ref. \cite{ToB06}. This suppression effect thus
introduces an ambiguity in the absolute cross sections that is
significant at the level of the differences being shown in Fig.
\ref{fig:ti54mix}. Such suppressions, of the cross sections
predicted using the shell-model spectroscopy, may themselves be, at
least in part, a manifestation of the use of TNA calculated in a
truncated shell-model space and that exclude a large number of
(small amplitude) cross-shell configurations. Based on the limited
measurements available to date, there is no indication that the
(missing) physics that drives the suppression of cross-section
strength has implications for the shape of the residue momentum
distribution. Additional, more accurate exclusive final state data
are needed to assess these expectations further.

\section{Summary}

We have discussed the momentum distributions of the heavy residues
after two-nucleon knockout reactions using $LS$-coupling. The main
factor determining the width of these momentum distributions is
shown to be the transferred total orbital angular momenta $L$ of the
two nucleons. We provide insight into the expected widths of
momentum distributions from the removal of the nucleon pair from
different configurations showing that information can be gained from
and upon the strengths of the theoretical two nucleon amplitudes and
the contributing $L$ they generate. The unambiguous observation of
effects associated with specific pairs of nucleon orbitals may
require transitions to final states that are relatively pure or
simple configurations. Some illustrative examples were presented and
discussed.

The conclusion of previous work - that the residue momentum
distribution was simply characteristic of the final state spin - has
been considered in further detail.  It is true that, generally,
higher spin final states will lead to wider residue momentum
distributions, but that the details of the shell model two-nucleon
overlap are important in understanding the details of the residue
momentum distributions. Quantitative testing and confirmation of
such sensitivity to the underlying structure will be essential for
the exploitation of two-nucleon knockout methods and their extension
for deformed nuclei.

The critical importance of configurations of different parity in
enhancing pairing correlations is demonstrated by consideration of
the angular correlations inherent in the two-nucleon density. Whilst
discussed here in the context of two-nucleon removal reactions and
enhancements of two-nucleon removal cross sections, such
considerations, of large basis TNA, and the importance of small
admixtures of different parity is entirely general. In the context
of the suppression of shell model strength, previous studies with
radioactive beams have demonstrated that the theoretical cross
sections overestimate experiment by about a factor of two.  It will
be important to experimentally verify the influence of cross-shell
excitations on structurally better-understood cases, such as for
\nuc{12}{C}, \nuc{16}{O} and \nuc{40}{Ca}, to clarify the extent to
which the necessary reductions may depend (in part) on the truncated
model spaces used. It will also be important to further assess the
importance of cross shell proton-excitations in studies of islands
of inversion using the two-proton knockout methodology (see e.g.
\cite{GAB07a,AAB08,FRM10}), where very strong reductions of
two-proton knockout cross sections are observed.

Here our emphasis has been on light and medium mass projectiles.
Another interesting example is the two-proton removal reaction from
\nuc{208}{Pb}; not only are there a large number of active orbitals,
producing a plethora of states, but the majority of states are good
two-proton hole configurations with minimal mixing.

The study of such reactions with odd-mass projectiles brings an
added layer of complication with, typically, each final state being
populated via several nucleon pair total angular momenta. The widths
of the residue momentum distributions are then no longer simply
related to a single final state spin.  However, the underlying
structure sensitivity discussed here may still yield characteristic
widths for different final states in the same residue, somewhat
independent of the final state spin.

\begin{acknowledgments}
This work was supported by the United Kingdom Science and Technology
Facilities Council (STFC) through Research Grant No. ST/F012012. ECS
gratefully acknowledges support from the United Kingdom Engineering
and Physical Sciences Research Council under Grant No. EP/P503892/1.
\end{acknowledgments}

\bibliographystyle{thesis}
%\bibliography{/Users/ecsimpson/refs/refs}

\end{document}